\documentclass[%
 reprint,
 amsmath,amssymb,
 aps,
 pre,
]{revtex4-1}

\usepackage{graphicx}
\usepackage{dcolumn}
\usepackage{bm,romannum}
\usepackage{epstopdf}

\begin{document}

\preprint{APS/Bhateja-2020}

\title{Velocity scaling in the region of orifice influence in silo draining under gravity}
\author{Ashish Bhateja}
\email{ashish@iitgoa.ac.in}
\affiliation{School of Mechanical Sciences, Indian Institute of Technology Goa, Ponda 403401, Goa, India.}

\date{\today}
\begin{abstract}
This study utilizes computations based on soft-particle discrete element method for investigating the scaling of velocity in a two-dimensional silo draining under gravity. We focus on the region situated directly above and in proximity to the outlet, referred to as the region of orifice influence (ROI). The velocity at the exit scales with the outlet size, in agreement with previous studies. The velocity in the ROI upstream of the outlet, however, does not collapse to a single curve when scaled with the outlet size. We show that the height of an \textit{equi-inertial} curve, which is defined to be a curve on which the inertial number is constant, can be employed for scaling the velocity in the ROI. The velocity corresponding to an equi-inertial curve, when measured relative to the outlet velocity, is considered for scaling. Results show that the scaling holds very well for low inertial number corresponding to the dense flow regime, whereas it breaks down for high inertial number region.
\end{abstract}

\maketitle


\section{Introduction}
Silos are commonly utilized for storing particulate materials and occupy an important place in several industrial applications. Typically, the silos are emptied by allowing the grains to exit under gravity through an opening \cite{nmbook,tighe2007,nedderman1982}. Granular materials draining under gravity through an outlet located at the base of a silo exhibit a variety of flow features, which have been the subject of intensive research for decades \cite{tighe2007,nedderman1982,brown,ahn2008,aguirre2010,aguirre2011,perge2012,janda2012,vidyapati2013,staron2014,rubio2015,peralta2017,to2017,madrid2019,
zuriguel2019}, and a general consensus on many issues is yet to be attained.

One of the primary concerns is the development of a robust mass flow rate correlation for the flow of granular media through an orifice \cite{beverloo1961,nedderman1982,mankoc2007,cleary2011,janda2012,benyamine2014,rubio2015,gella2017}. A widely-known mass flow rate correlation was given by \citet{beverloo1961}, which may also be obtained by utilizing the dimensional analysis. Beverloo expression has been tested in a variety of situations where the silo discharge is in the continuous flow regime \cite{martinez2008,perge2012,staron2012,cordero2015}. Later, an exponential correction was proposed by \citet{mankoc2007}, catering to a wide spectrum of orifice size. An experimental study of \citet{janda2012} in a two-dimensional silo, aiming to derive an expression for the mass flow rate, reported the appearance of self-similar velocity and density profiles at the outlet in the monodisperse granular flow comprising 1 mm diameter stainless steel beads. An important finding was the existence of the scaling of velocity at the exit while considering the outlet size ($D$) as a length scale. This scaling relates well to the notion of free fall arch, the existence of which is recently questioned by \citet{rubio2015}.

Gella and co-workers \cite{gella2017} also experimentally obtained the self-similar velocity profiles and the scaling of velocity with $D$, at the outlet, in the monodisperse granular flow with larger particle size in a quasi two-dimensional silo, in line with what \citet{janda2012} reported. The self-similar velocity profiles have also been recovered numerically by \citet{zhou2015} using the discrete particle computations for monodisperse and bidisperse granular media in a planar silo. Recently, an experimental investigation of \citet{fullard2018} presented the scaling of velocity with the flow rate in the steady and transient states in a three-dimensional gravity-induced silo discharge. Their analysis, utilizing Particle Image Velocimetry, was quasi two-dimensional as the scaling was based on the velocities of particles accessed through one side-wall of the silo. A flow-rate dependent velocity scaling is reported at locations situated away from the orifice in the region of small inertial number \cite{dacruz2005,jop2006}.

In light of the above and considering substantial variation in flow characteristics as grains approach the outlet \cite{staron2012,vidyapati2013}, a question which arises is whether orifice size, $D$, can scale the velocities in the region situated directly above and close to the outlet. The objective of this work is, therefore, to examine velocity scaling in proximity to the orifice in a two-dimensional silo draining under gravity, employing computations based on soft-particle discrete element method \cite{cundall1979,bkm2003}. The paper is organized as follows. Section~\ref{sec:DEM} details the computational methodology and simulation set-up. Results are provided in Sec. \ref{sec:RD}, followed by concluding remarks in Sec.~\ref{sec:conclude}.

\section{Computational details}
\label{sec:DEM}

\begin{figure}[ht!]
\begin{center}
\includegraphics[scale=0.55]{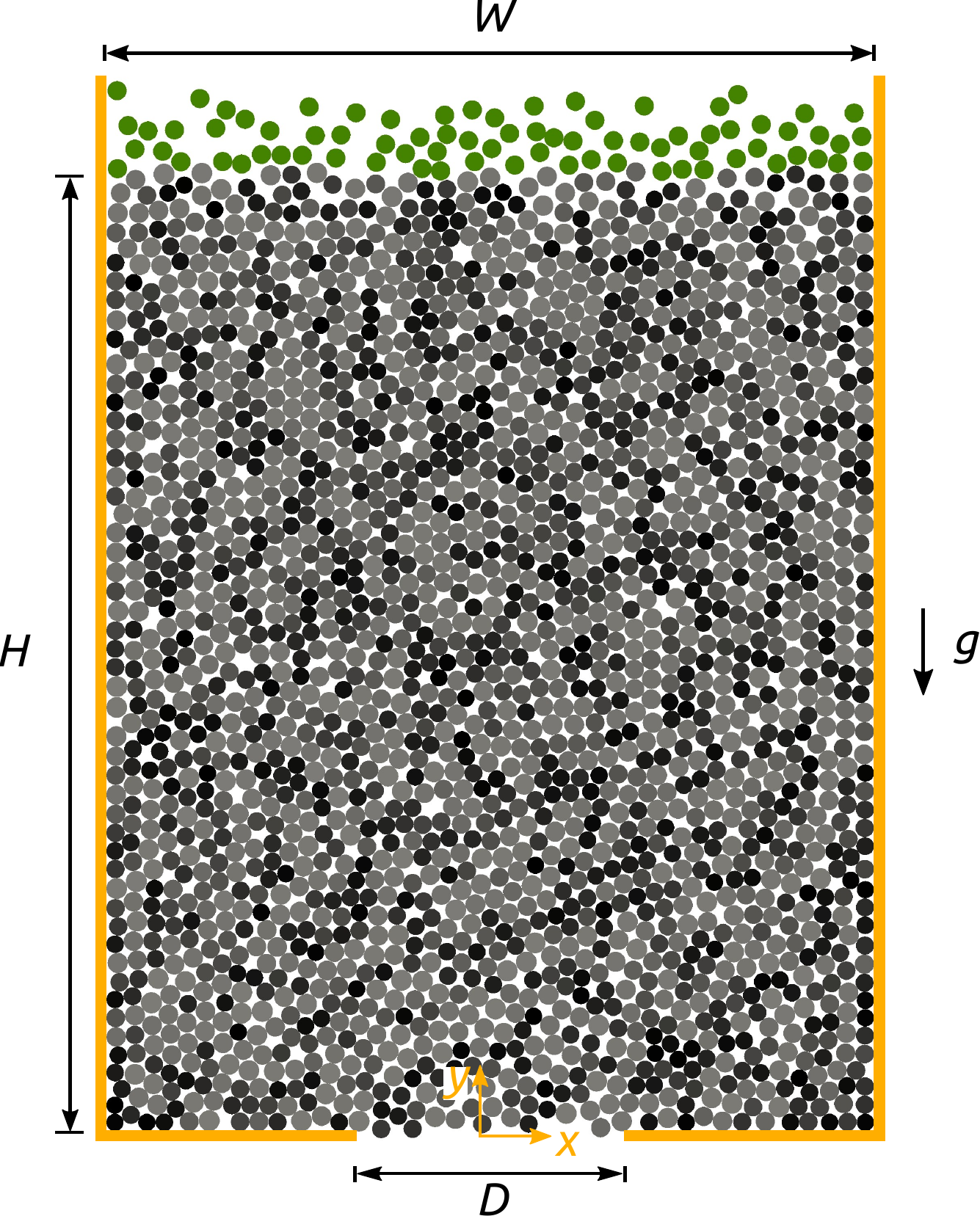}
\caption{A simulation snapshot of granular discharge in a two-dimensional silo for initial fill height $H=50d$; the fill heights up to seven times of it are considered in this study as mentioned in the text. Coordinate axes and direction of gravitational acceleration $g$ are indicated appropriately. The reinserted grains are shown in green color at the top.}
\label{fig:ss}
\end{center}
\end{figure}

Computations are carried out by utilizing the soft-particle discrete element method \cite{cundall1979,bkm2003} for simulating the granular flow in a two-dimensional silo; a typical snapshot is displayed in Fig. \ref{fig:ss}. The silo operates in the continuous-discharge mode in which the exited grains are reinserted into the system at random horizontal locations above the top layer with zero velocity, thereby maintaining a constant fill height. This set-up achieves a steady state for longer duration, providing a large statistics for averaging. 

The grains are considered to be dry, non-cohesive and deformable disks of density $\rho$ and mean diameter $d$, with a size polydispersity of $\pm 10 \%$ so as to avoid crystallization and its possible consequences on the flow \citep{potapov1996}. The disks are inelastic and frictional. The interaction between the grains is modelled through linear spring-dashpot force scheme, with an impact velocity-independent restitution coefficient \citep{cundall1979,zhang1996,bkm2003,shafer1996}. The friction during contact between the grains is incorporated by employing the Coulomb friction criterion \citep{shafer1996}. The same force scheme is employed for interaction between the walls and grains; the details of the force model are provided elsewhere \cite{bhateja2016,bhateja2020}.

Six different orifice size $D$ are used, varying between $9d$ and $14d$ in increments of one mean particle diameter, $d$. The outlet size is taken to be larger than $6d$ so as to avoid jamming due to the formation of a stable arch across the orifice \citep{mankoc2007,janda2008,kondic2014}. The silo width $W$ is kept fixed at $40d$, ensuring no effect of the side-walls on the flow adjacent to the outlet for the chosen range of $D$, i.e., $W > 2.5 D$ \citep{nedderman1982,vidyapati2013}. Fill height $H$ of the grains is varied between $50d$ and $350d$ in steps of $50d$. The number of particles $N$ corresponding to the fill height are given in the format [$H:N$] as follows: [$50d:2150$]; [$100d:4300$]; [$150d:6505$]; [$200d:8700$]; [$250d:10920$]; [$300d:13160$] and [$350d:15345$].

All quantities of interest are made non-dimensional with mean diameter $d$, density $\rho$ and gravitational acceleration $g$. The simulations are carried out with normal spring stiffness $k_n=10^6$, whereas no spring is considered for tangential direction, i.e., tangential spring stiffness $k_t=0$. The restitution and friction coefficients between the contacting grains are $e_p = 0.9$ and $\mu_p = 0.4$, respectively. The same values are used for wall-grain interactions, i.e., $e_w = 0.9$ and $\mu_w = 0.4$. The equations of motion are integrated by employing the velocity-Verlet algorithm \cite{kruggel2008b,allen1989} with an integration time step equal to $\delta t=10^{-4}$.

The data reported in this study are recorded after achieving the steady state and averaged over 50 simulation runs, with every simulation begins with a different initial configuration and runs for one million time steps. The averaging scheme employs coarse-graining technique \citep{goldhirsch2010,weinhart2013,artoni2015}, utilizing a Heaviside step function with coarse-grained width $w$ equal to mean particle diameter $d$. The $x$ and $y$ components of mean translational velocity vector $\bm{v}$ at the center of a bin are calculated as, respectively,
\begin{eqnarray}
v_x &=& \frac {1}{N_b} \sum_{i=1}^{N_b} u_{xi},\\
v_y &=& \frac {1}{N_b} \sum_{i=1}^{N_b} u_{yi},
\end{eqnarray}
where $u_{xi}$ and $u_{yi}$ are the components of instantaneous velocity vector $\bm{u}$ of particle $i$ along $x$ and $y$ directions, respectively, and $N_b$ is the number of particles lying within $w/2$ distance from the bin center such that $|x_i-x_b|\leq w/2$ and $|y_i-y_b|\leq w/2$, and $\bm{r}_i$=($x_i$, $y_i$) and $\bm{r}_b$=($x_b$, $y_b$) are the coordinates of particle $i$ and bin center, respectively. The total stress tensor $\bm{\sigma} = \bm{\sigma}_c + \bm{\sigma}_s$ is computed considering the collisional ($\bm{\sigma}_c$) and streaming ($\bm{\sigma}_s$) stress contributions \cite{tripathi2010,bhateja2020}. The streaming stress component is calculated by ensuring that the fluctuation velocity is independent of the bin size, following \citet{artoni2015}.

\section{Results and discussion}
\label{sec:RD}
\subsection{Region of Orifice Influence}
\label{sec:RD1}
\begin{figure}[ht!]
\begin{center}
\includegraphics[scale=0.6]{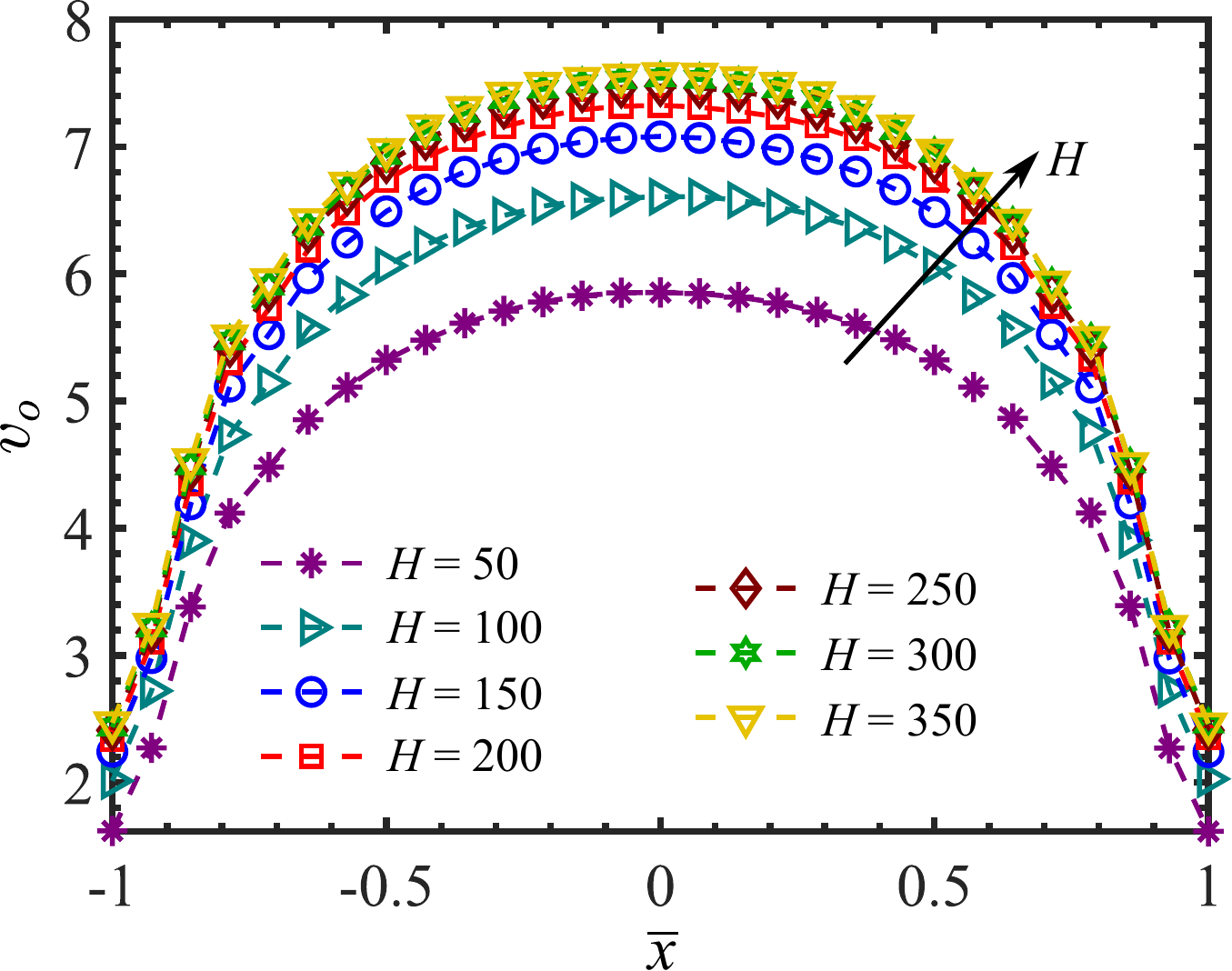}
\caption{Horizontal profiles of the magnitude of mean velocity at the outlet, $v_o$, for different fill heights for $D=14$. Error bars, based on the standard error \cite{altman2005}, are not shown due to their size smaller than the symbols.}
\label{fig:velH}
\end{center}
\end{figure}

We first present the variation of the magnitude of mean velocity at the outlet, $v_o$, with $\overline{x}$ for all fill heights for $D=14$ in Fig.~\ref{fig:velH}, where $\overline{x}=2x/D$. The results obtained are qualitatively similar for all orifice sizes. We observe in Fig.~\ref{fig:velH} that, for a given $H$, the velocity $v_o$ increases towards the center and attains a maximum value, which is consistent with the reported experimental findings \cite{janda2012,gella2017}. Also, at a fixed $\overline{x}$, $v_o$ rises as $H$ increases and becomes nearly constant beyond $H=200$. Accordingly, our subsequent discussion is based on the system with $H=250$.

\begin{figure}[ht!]
\begin{center}
\includegraphics[scale=0.55]{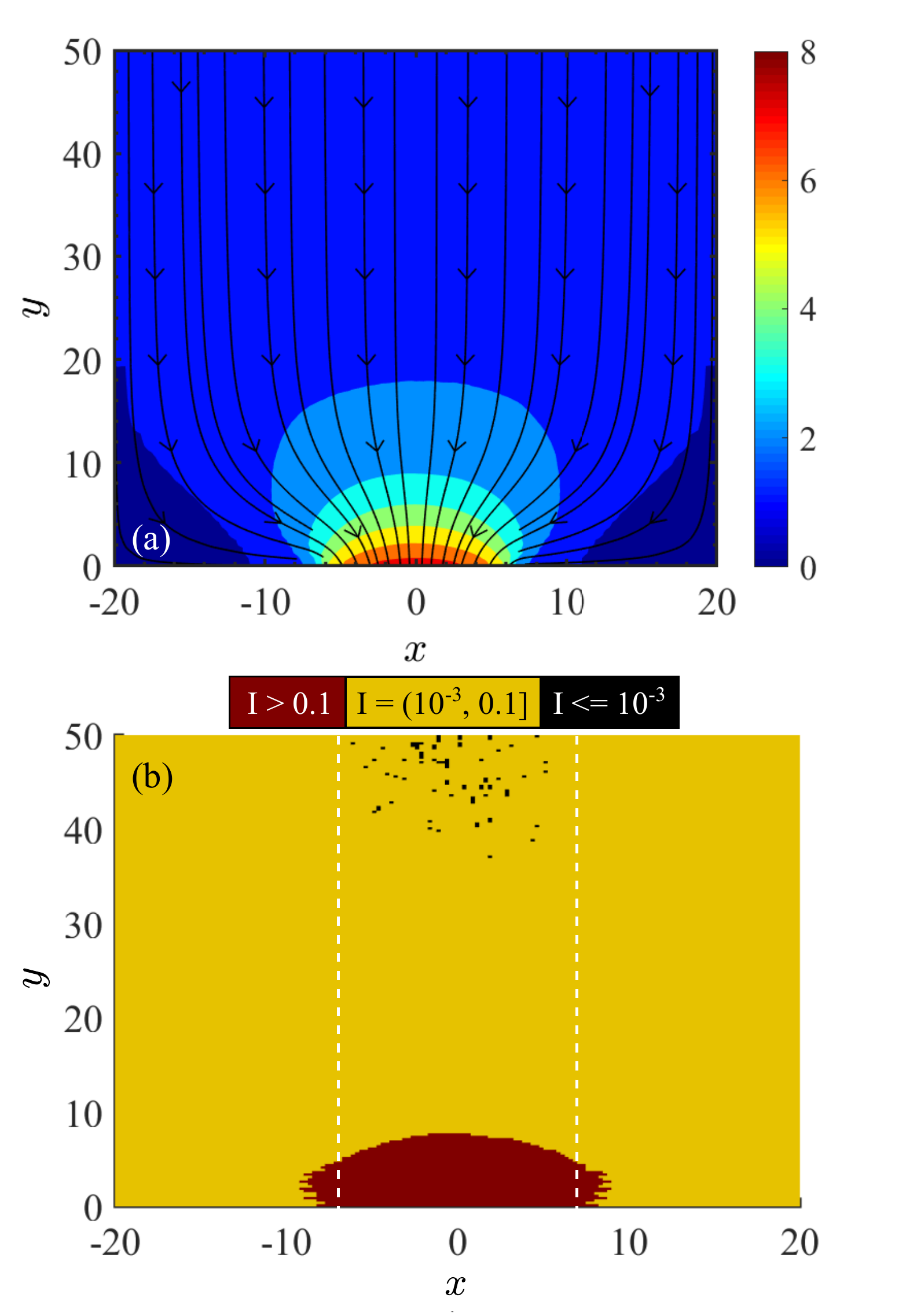}
\caption{(a) Spatial map of the magnitude of mean velocity $v$, and streamlines. (b) Spatial distribution of inertial number $I$. A few points shown near the top correspond to $I \leq 10^{-3}$ , denoting the quasi-static flow regime. Color scales are provided appropriately for each distribution. The data are considered for $D=14$ and $H=250$, while displayed for clarity upto $y=50$.}
\label{fig:vel_inum}
\end{center}
\end{figure}

As mentioned in the Introduction, we intend to study granular flow adjacent to the orifice. It is, therefore, essential to explore the region wherein the orifice significantly affects the flow. To this end, we examine spatial distribution of mean velocity $v$ for $D=14$ in Fig.~\ref{fig:vel_inum}(a). As expected, the velocity is larger in the region located upstream of the orifice in its vicinity and lowest adjacent to the base and side walls, consistent with earlier investigations \cite{staron2012, staron2014}. The velocity becomes roughly constant upstream of the flow and varies significantly adjacent to the outlet, which is clearly demonstrated by the streamlines that are nearly straight far away from the orifice and converge into the outlet. In order to further highlight the flow condition near the orifice, Fig.~\ref{fig:vel_inum}(b) presents the spatial distribution of inertial number \cite{midi2004,dacruz2005,jop2006} $I=\dot{\gamma}d/\sqrt{P/\rho}$, where $P$ is the pressure given by the trace of stress tensor $\bm{\sigma}$, and $\dot{\gamma} = (2\bm{D:D})^{1/2}$ is the shear rate with $\bm{D} = (\bm{G_s} + \bm{G_s}^T)/2$ being the symmetric part of the traceless velocity gradient tensor $\bm{G_s}=[\nabla{\bm{v}} - (\nabla\cdot\bm{v})\,\bm{I}/2]$. The dense ($I=(10^{-3},0.1]$) and rapid ($I>0.1$) flow regimes are shown corresponding to the range of inertial number given by \citet{dacruz2005}. The rapid flow regime occurs in proximity to the orifice and largely confined within the central domain, the region enclosed between the dashed lines shown in Fig.~\ref{fig:vel_inum}(b). 

In light of the above, the region wherein the effect of the orifice is predominant is labelled as \textit{Region of Orifice Influence} (ROI), spreading largely within the central domain and located directly above the outlet. The vertical extent of the ROI from the orifice may be estimated as the height beyond which mean velocity becomes almost constant, i.e., streamlines become straight. Hereafter, we focus our discussion on the flow in the ROI. Note that, for a given orifice size, the left and right boundaries of the central domain correspond to $\overline{x}=-1$ and $1$, respectively.

%

\subsection{Velocity scaling}

\begin{figure}[ht!]
\begin{center}
\includegraphics[scale=0.55]{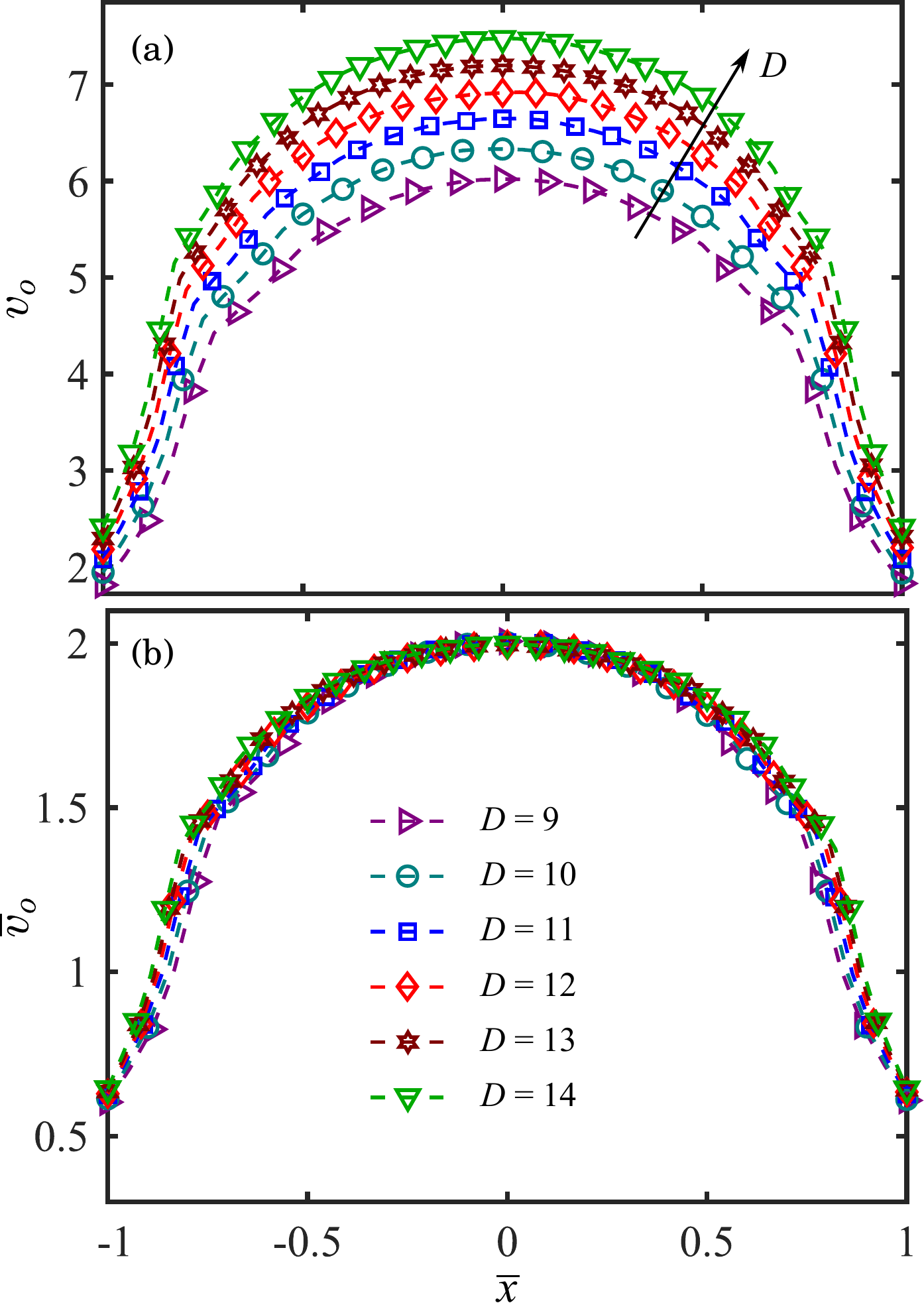}
\caption{Variation of (a) $v_o$ and (b) $\overline{v}_o$  with $\overline{x}$ for different outlet size $D$. Legend for both plots is given in (b). Error bars are not displayed due to their size smaller than the symbols.}
\label{fig:velscaleD}
\end{center}
\end{figure}

We first examine the scaling of $v_o$ considering $D$ as a length scale. Fig.~\ref{fig:velscaleD}(a) shows the variation of $v_o$ with $\overline{x}$ for different $D$. Expectedly, at a given $\overline{x}$, $v_o$ increases as the orifice widens. For a fixed orifice size, the velocity rises with $\overline{x}$ and attains a maximum value at the center. The variation of scaled velocity $\overline{v}_o$ with $\overline{x}$ is presented in Fig.~\ref{fig:velscaleD}(b), where $\overline{v}_o= v_o/(gD)^{1/2}$. The velocity profiles are self-similar as the data corresponding to different $D$ collapse onto a single curve, in notable agreement with what is reported in experiments \cite{janda2012,gella2017} and simulations \cite{zhou2015}.

%

\begin{figure}[ht!]
\begin{center}
\includegraphics[scale=0.55]{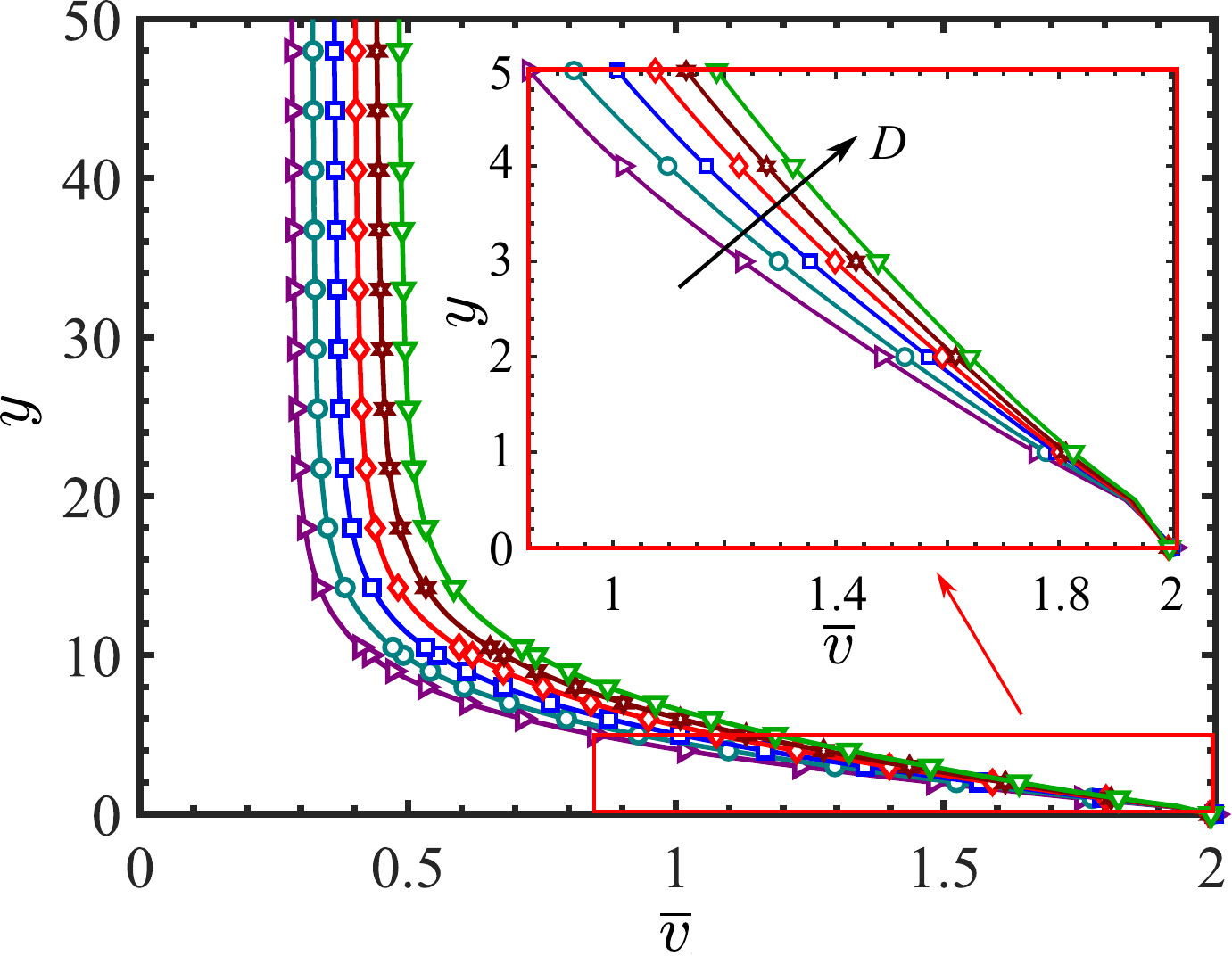}
\caption{Variation of scaled velocity $\overline{v}$ with $y$ for different $D$ at $\overline{x}=0$. Legend is provided in Fig.~\ref{fig:velscaleD}(b). Inset shows the variation of $\overline{v}$, at $\overline{x}=0$, with $y$ for a small region spanning upto $5$ particle diameter. Error bars are smaller than the symbols, thus, not shown.}
\label{fig:scaledvelx0}
\end{center}
\end{figure}

Let us further examine whether $D$ can serve as a length scale for scaling the velocities in the region upstream of the orifice. To this end, for the sake of simplicity and without loss of generality, we consider variation of scaled velocity $\overline{v}$ along vertical direction $y$ at $\overline{x}=0$ as presented in Fig.~\ref{fig:scaledvelx0}, where $\overline{v}=v/(gD)^{1/2}$. The data for $\overline{v}$ corresponding to different $D$ lie on distinct curves as we move away from the orifice center along $y$ direction. In fact, as the inset in Fig.~\ref{fig:scaledvelx0} displays, the scaling breaks down within the height of one particle diameter. Thus, in the absence of velocity scaling with $D$, a natural question one could ask about the existence of a parameter capable of scaling the velocities in the ROI.

\begin{figure}[ht!]
\begin{center}
\includegraphics[scale=0.58]{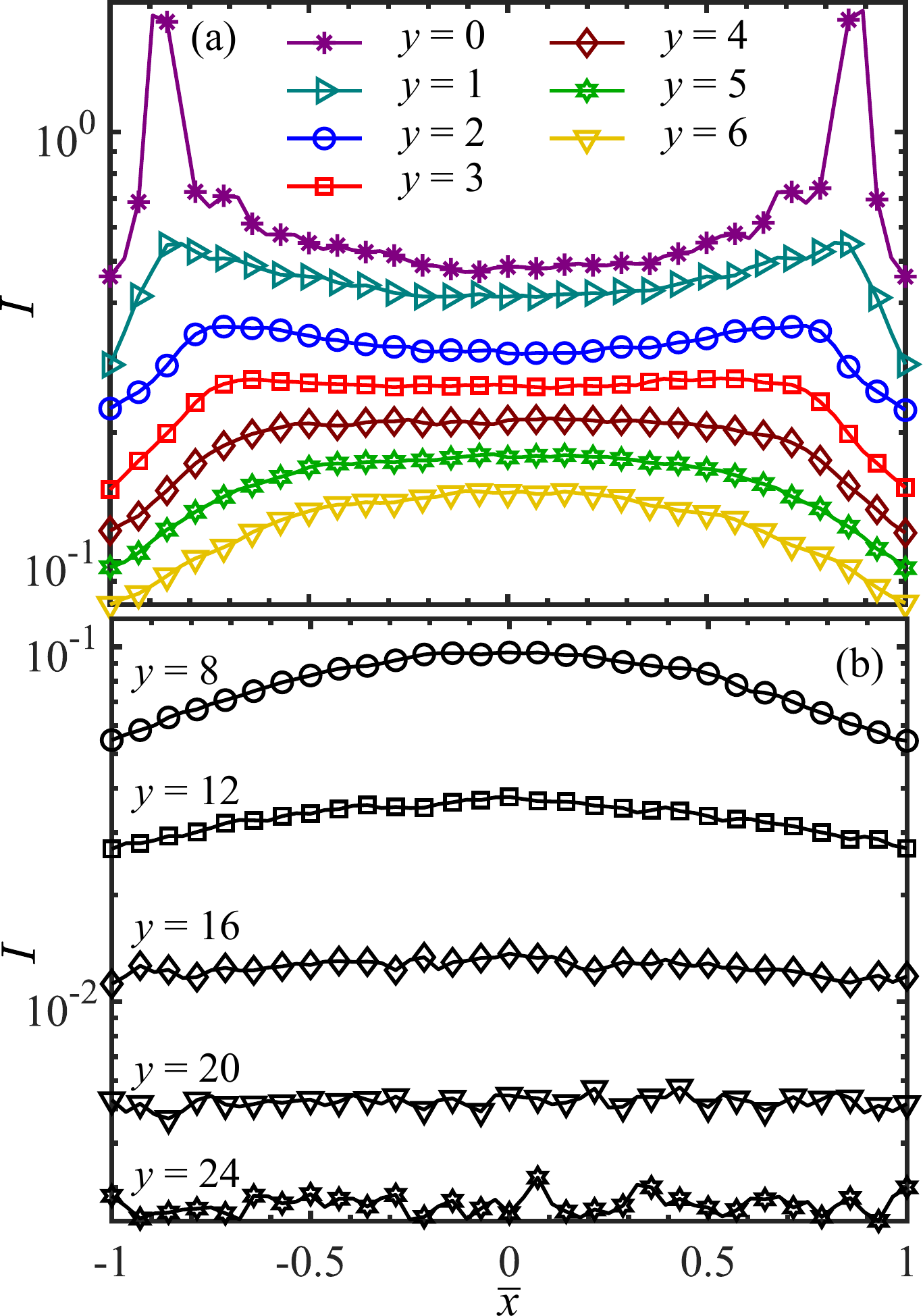}
\caption{Variation of inertial number $I$ with $\overline{x}$ in the central domain for different vertical locations $y$. Error bars are not displayed due to their size smaller than the symbols.}
\label{fig:inum}
\end{center}
\end{figure}

One such parameter may be the vertical location, $h_e$, of an \textit{equi-inertial} curve, which may be defined as a curve corresponding to a constant inertial number. The height $h_e$ is measured vertically from the outlet, and may have a dependence on $\overline{x}$, i.e., $h_e=h_e(\overline{x})$, given the variation of inertial number $I$ with $\overline{x}$ in Figs.~\ref{fig:inum}(a) and \ref{fig:inum}(b) for several vertical locations. The inertial number varies considerably with $\overline{x}$ at the locations ($y$) close to the outlet, whereas $I$ is nearly independent of $\overline{x}$ in the region far away from the outlet, similar to the case of mean velocity and streamlines in Fig.~\ref{fig:vel_inum}(a). 

We further notice in Fig.~\ref{fig:inum} that $I$ decreases monotonically as we travel upstream of the flow. For the sake of simplicity and without loss of generality, we consider a few values of $I$ in order to locate the equi-inertial curves, and accordingly velocity scaling will be examined in the ROI. Here, we consider the following ten values of $I=\{0.0015,0.0025,0.005,0.01,0.05,0.1,0.15,0.2,0.25,0.3\}$. For brevity, the results are presented for $I=\{0.0015,0.1,0.3\}$. These three $I$ values are chosen so as to focus on three different flow regimes based on the range of inertial number \cite{dacruz2005}, as discussed in Sec.~\ref{sec:RD1}, where $I=0.3$ and $0.1$ correspond to the rapid flow and dense flow regimes, respectively, and $I=0.0015$ lies close to the quasi-static regime. From now onwards, the argument of $h_e$ is omitted for brevity.

\begin{figure*}[ht!]
\begin{center}
\includegraphics[scale=0.4]{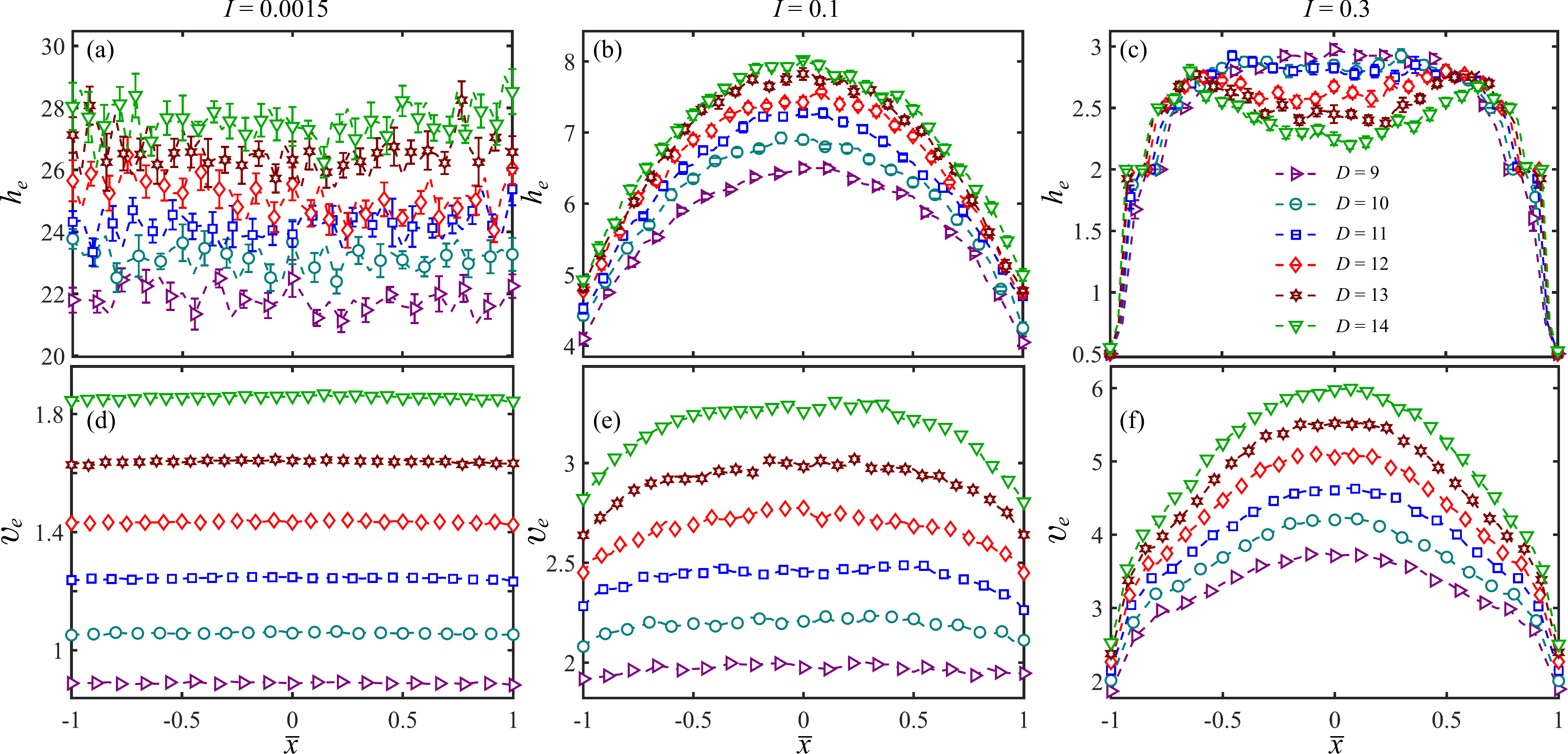}
\caption{(a)-(c) Variation of the vertical location of an equi-inertial curve $h_e$ with $\overline{x}$. Error bars are given by the standard error \cite{altman2005}. (d)-(f) Variation of $v_e$ with $\overline{x}$. Error bars are not displayed as these are smaller than the symbols. Data are presented for different $D$ and the legend for all plots is provided in (c).}
\label{fig:heve}
\end{center}
\end{figure*}

Figs.~\ref{fig:heve}(a)-\ref{fig:heve}(c) show the variation of $h_e$ with $\overline{x}$ for three $I$ values and six outlet sizes. As evidenced in Fig.~\ref{fig:heve}(a), for a given $D$, $h_e$ is roughly constant in the ROI for $I=0.0015$, whereas it varies significantly with $\overline{x}$ for $I=0.1$ and reaches a maximum value in the center as shown in Fig.~\ref{fig:heve}(b). Moreover, $h_e$ increases with $D$ at a given $\overline{x}$. The mean velocity at an equi-inertial curve, $v_e$, is obtained and its variation with $\overline{x}$ for three $I$ values for different $D$ are shown in Figs.~\ref{fig:heve}(d)-\ref{fig:heve}(f). For all three scenarios, at a given $\overline{x}$, $v_e$ rises as the orifice widens. The behaviour of $v_e$ relates well with $h_e$ for $I=0.0015$ and $0.1$, however, the situation is quite opposite for $I=0.3$ (cf. Figs.~\ref{fig:heve}(c) and \ref{fig:heve}(f)).

\begin{figure*}[ht!]
\begin{center}
\includegraphics[scale=0.4]{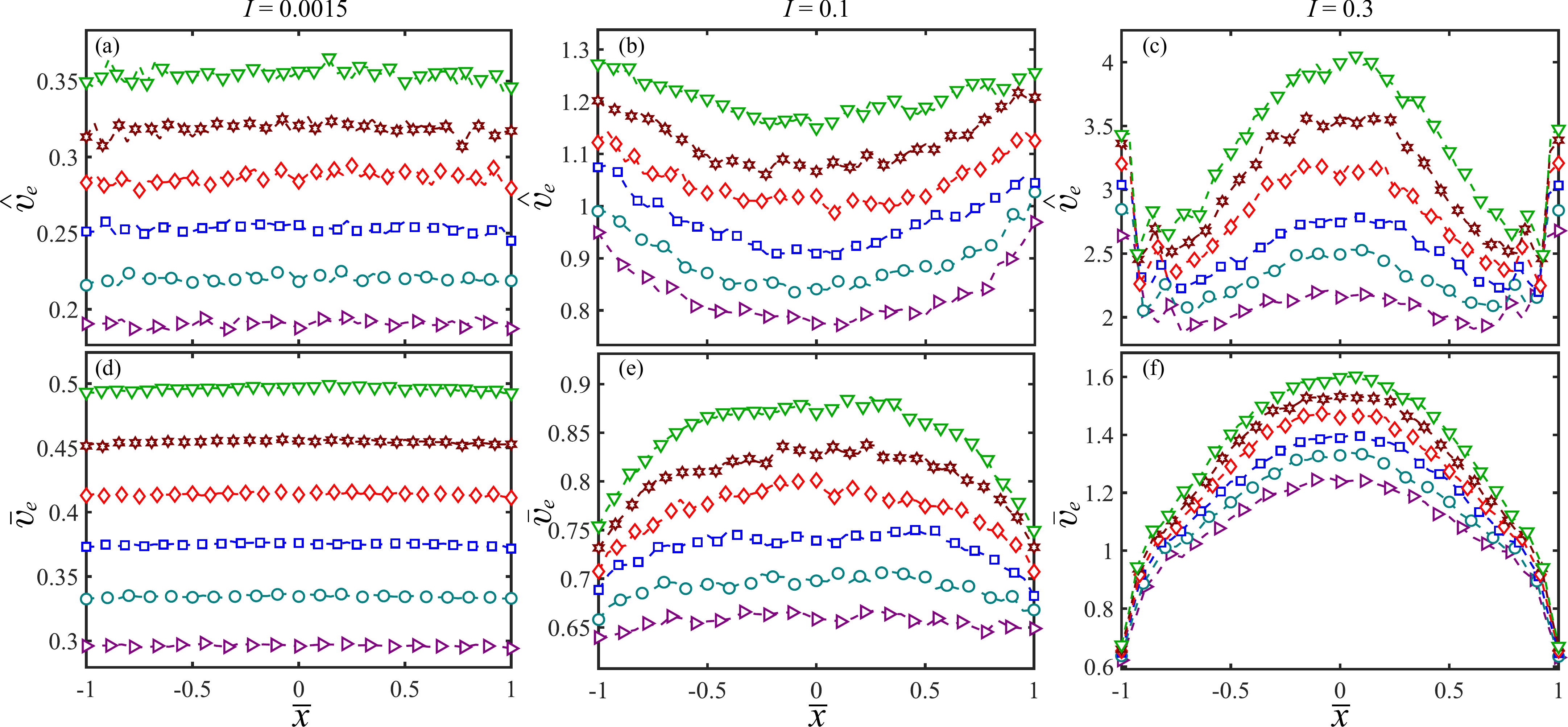}
\caption{(a)-(c) Variation of the scaled velocity $\hat{v}_e=v_e/(gh_e)^{1/2}$ with $\overline{x}$. (d)-(f) Variation of the scaled velocity $\overline{v}_e=v_e/(gD)^{1/2}$ with $\overline{x}$. Legend for all plots is given in Fig.~\ref{fig:heve}(c). Error bars are not shown due to their size smaller than the symbols.}
\label{fig:vcScale}
\end{center}
\end{figure*}

We now examine the scaling of $v_e$ with $h_e$ and $D$. The variation of scaled velocities, $\hat{v}_e$ and $\overline{v}_e$, with $\overline{x}$ for three $I$ values and six outlet sizes are shown in Fig.~\ref{fig:vcScale}, where $\hat{v}_e=v_e/(gh_e)^{1/2}$ and $\overline{v}_e=v_e/(gD)^{1/2}$. It is clear that neither $h_e$ nor $D$ scale $v_e$, given the absence of data collapse onto a single curve for different orifice sizes. The latter is expected in light of what is observed in Fig.~\ref{fig:scaledvelx0}, and the former signifies that the grains do not fall freely under gravity in the region of orifice influence, otherwise, $v_e$ would possibly have scaled with $h_e$ according to purely kinematic arguments. It is noteworthy, in passing, that even though the behaviours of $h_e$ and $v_e$ are quite opposite for $I=0.3$, the variation of $\hat{v}_e$ is similar for all six outlet sizes, as shown in Fig.~\ref{fig:vcScale}(c). In the absence of obtaining the scaling of $v_e$ with $h_e$, we examine next the scaling of relative velocity $v_r=v_e-v_o$, i.e., the velocity at an equi-inertial curve with reference to the velocity at the outlet $v_o$, with $h_e$ and $D$.

%
\begin{figure}[ht!]
\begin{center}
\includegraphics[scale=0.58]{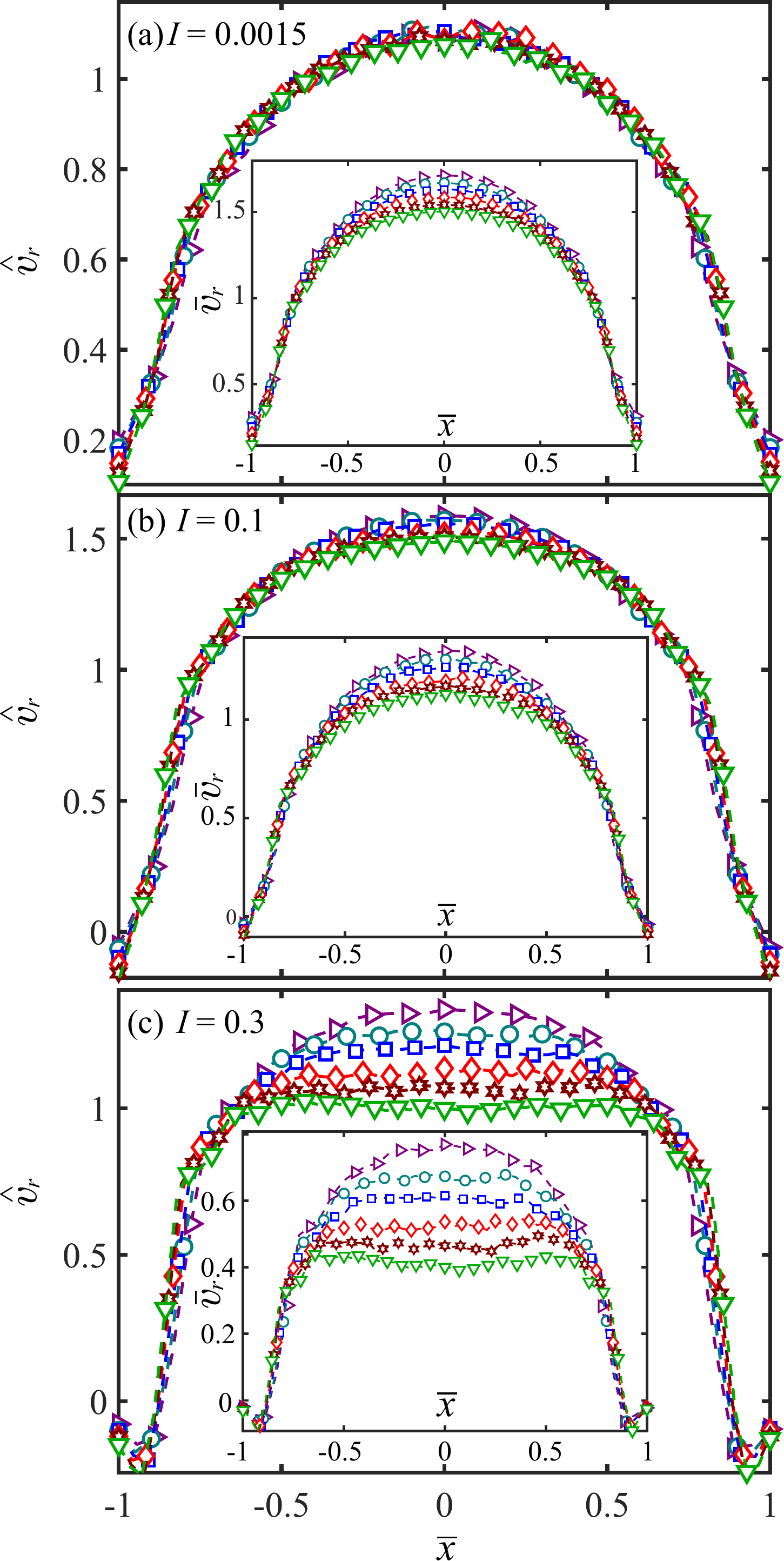}
\caption{(a)-(c) Scaled relative velocity $\hat{v}_r=v_r/(gh_e)^{1/2}$ with $\overline{x}$ for three $I$ values and six orifice sizes. Variation of $\overline{v}_r=v_r/(gD)^{1/2}$ with $\overline{x}$ is shown in the insets. Error bars are not displayed as these are smaller than the symbols. Legend for all plots is given in Fig.~\ref{fig:heve}(c).}
\label{fig:vrhe}
\end{center}
\end{figure}

\begin{figure}[ht!]
\begin{center}
\includegraphics[scale=0.58]{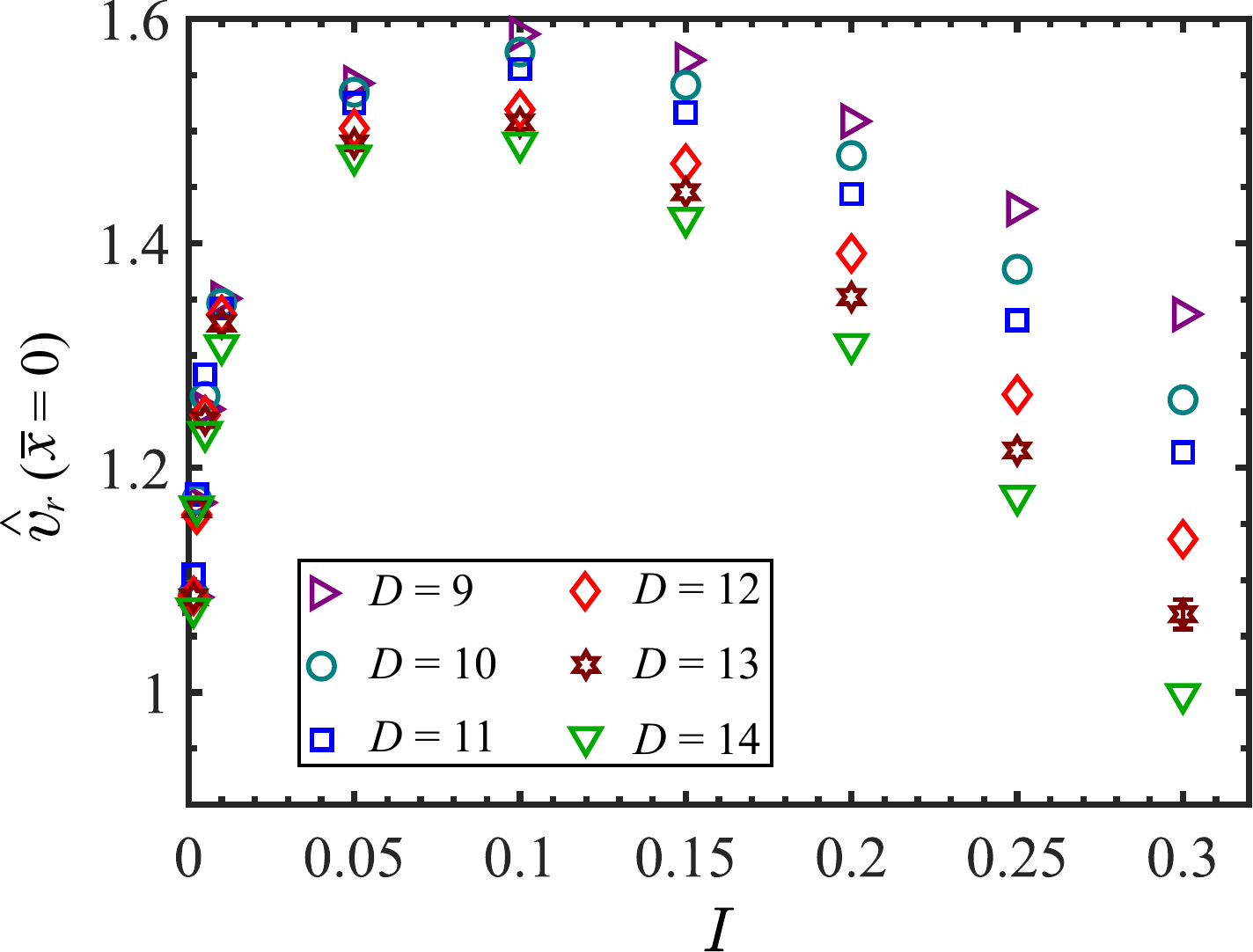}
\caption{Variation of scaled relative velocity $\hat{v}_r$, at $\overline{x}=0$, with inertial number $I$ for different $D$. Error bars which are smaller than the symbols are not shown for clarity.}
\label{fig:vrx0}
\end{center}
\end{figure}

Fig.~\ref{fig:vrhe} shows the variation of $\hat{v}_r$ with $\overline{x}$ for three $I$ values and different outlet sizes, where $\hat{v}_r=|v_r|/(gh_e)^{1/2}$ and $|\cdot|$ denotes the absolute value. The data collapse very well for $I=0.0015$, and the scaling tends to break down for $I=0.3$. This observation indicates that the scaling holds well for equi-inertial curves corresponding to low $I$ values. In order to highlight this  without loss of generality, the variation of $\hat{v}_r$ at $\overline{x}=0$ with $I$ for different $D$ is displayed in Fig.~\ref{fig:vrx0}, which demonstrates that $h_e$ scales $v_r$ very well for low $I$ values. Further, it is worth emphasizing that the collapse is, nevertheless, non-trivial in light of the dissimilar shapes of velocity profiles at the outlet and equi-inertial curve (cf. Figs.~\ref{fig:velscaleD}(a) and \ref{fig:heve}(d)). Graphs displaying the velocity scaling for other $I$ values are provided in the Supplementary Material \cite{supmat}. It is important to note that such a good collapse is not obtained considering $D$ as a length scale for the scaling of relative velocity, which is demonstrated in the insets of Fig.~\ref{fig:vrhe} displaying the variation of $\overline{v}_r$ with $\overline{x}$ for various orifice sizes, where $\overline{v}_r=|v_r|/(gD)^{1/2}$. Further, the disparity observed in the curves in the middle portion of the central domain increases for higher inertial number, similar to what occurs for $\hat{v}_r$ (see Supplementary Material \cite{supmat}).

%
\section{Conclusions}
\label{sec:conclude}
This study presents a detailed analysis based on soft-particle discrete element computations for investigating the scaling of velocity in the region of orifice influence (ROI), located directly above and in the neighbourhood of the outlet in a two-dimensional silo discharging under gravity. The velocity at the outlet, $v_o$, scales with the orifice size, in striking agreement with the reported experiments \cite{janda2012,gella2017} and discrete particle computations \cite{zhou2015}. However, no such scaling is obtained for the velocities in the region of orifice influence, while moving upstream of the orifice.

A possible choice for scaling the velocities in the ROI is the vertical location ($h_e$), measured from the orifice, of an equi-inertial curve on which the inertial number is constant. While considering, separately, $h_e$ and the outlet size for the scaling, the velocities corresponding to equi-inertial curves, $v_e$, for all six outlets do not collapse onto a single curve. The relative velocity $v_r$, which results when $v_e$ is measured with reference to $v_o$, scales very well with $h_e$ for $v_e$ corresponding to low inertial number belonging to the dense flow regime. However, no such scaling is obtained for $v_e$ corresponding to the equi-inertial curves for high inertial number in the region located close to the outlet. Further, $v_r$ does not scale well with the outlet size, suggesting the relevance of $h_e$ for the scaling of velocities in the region of orifice influence.

\section*{Acknowledgements}
I thank Prof. Neeraj Kumbhakarna for providing access to his computing facility for running simulations presented in this paper. I am grateful to Professor Devang Khakhar and Dr. Vaibhav Wasnik for insightful discussions.

\bibliography{mybib.bib}	
\section*{\cite{supmat} Supplementary material}
\begin{figure*}[ht!]
\begin{center}
\includegraphics[scale=0.55]{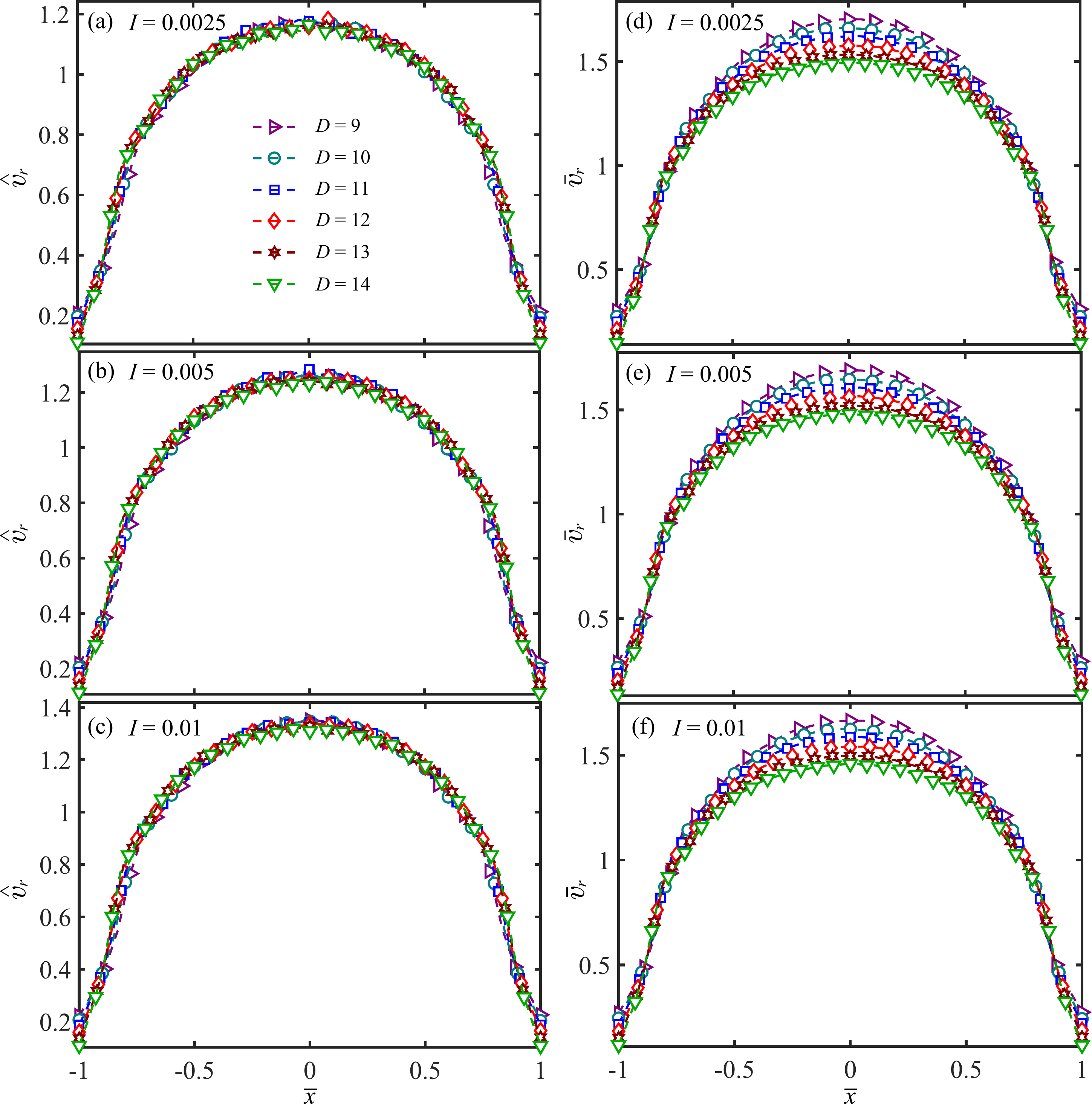}
\caption{(a)-(c) Variation of relative scaled velocity $\hat{v}_r$ with $\overline{x}$ for three $I$ values. (d)-(f) Variation of $\overline{v}_r$ with $\overline{x}$ for three $I$ values. Data are presented for six different outlets. Legend for all plots is given in (a). Error bars are not displayed due to their size smaller than symbols.}
\label{fig:sm1}
\end{center}
\end{figure*}

\begin{figure*}[ht!]
\begin{center}
\includegraphics[scale=0.55]{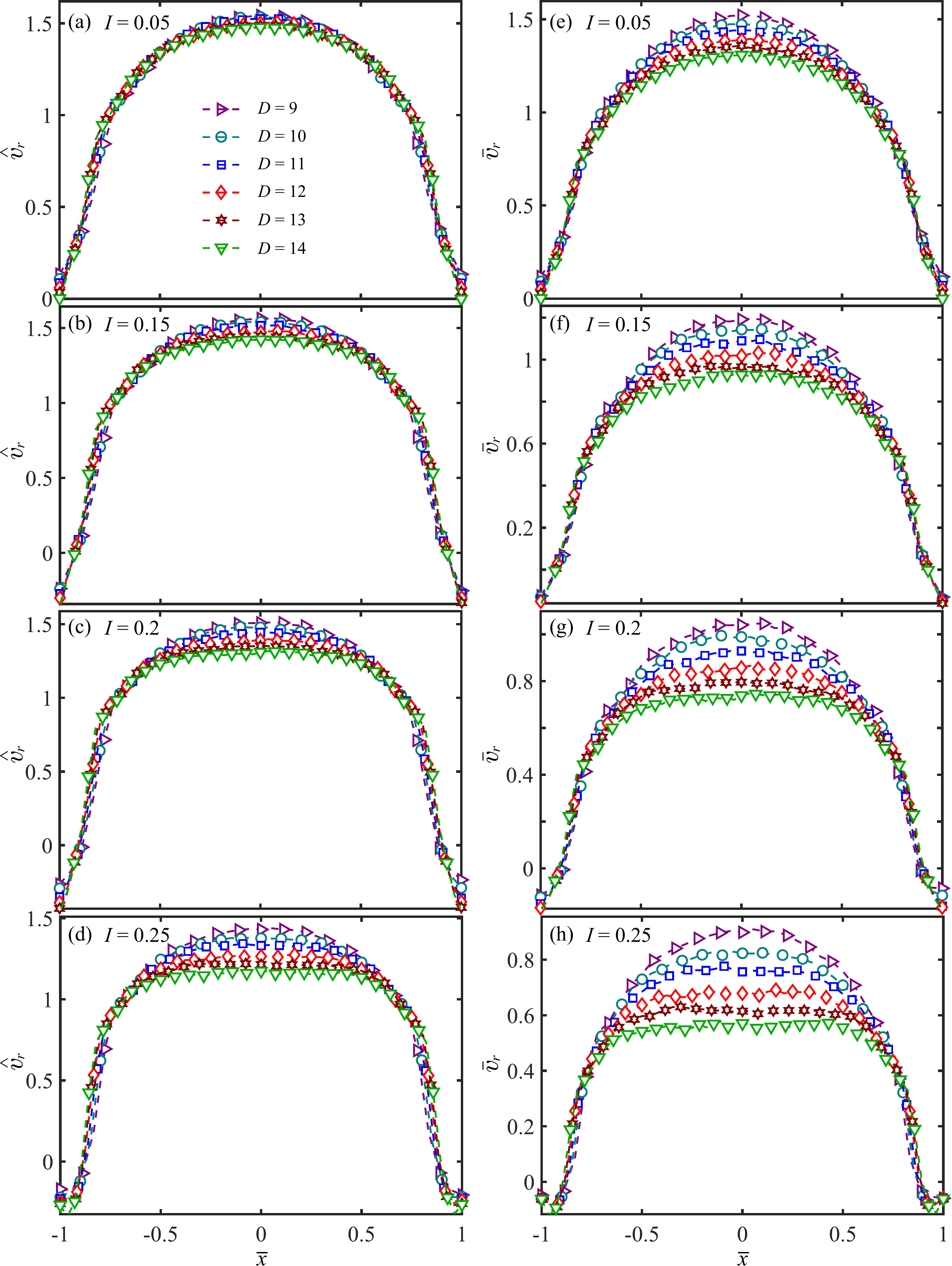}
\caption{(a)-(d) Variation of relative scaled velocity $\hat{v}_r$ with $\overline{x}$ for four different $I$ values. (e)-(h) Variation of $\overline{v}_r$ with $\overline{x}$ for four different $I$ values. Data are presented for six different outlets. Legend for all plots is given in (a). Error bars are not shown due to their size smaller than symbols.}
\label{fig:sm2}
\end{center}
\end{figure*}

[ht!]
\noindent In the main text, we show scaling of the relative velocity $v_r$ with $h_e$ and outlet size $D$ for three different values of inertial number ($I$). Here, Figs.~\ref{fig:sm1} and \ref{fig:sm2} show the scaled velocities $\hat{v}_r=v_r/(gh_e)^{1/2}$ and $\overline{v}_r=v_r/(gD)^{1/2}$ for seven different values of inertial number, i.e., \\$I=\{0.0025,0.005,0.01,0.05,0.15,0.2,0.25\}$. \\


\end{document}